# What is the function of inter-hemispheric inhibition?

Richard G. Carson[a,b,c]

[a]Trinity College Institute of Neuroscience and School of Psychology, Trinity College Dublin, Dublin 2, Ireland

[b]School of Psychology, Queen's University Belfast, Belfast, Northern Ireland, BT7 1NN, UK.

[c]School of Human Movement and Nutrition Sciences, The University of Queensland, QLD 4072, Australia

Correspondence: Richard G. Carson

Trinity College Institute of Neuroscience and School of Psychology

Trinity College Dublin

Dublin 2,

Ireland,

E-mail: richard.carson@tcd.ie

**Abstract**

It is widely supposed that following unilateral brain injury, there arises an asymmetry in inter-hemispheric inhibition which has an adverse influence upon motor control. I argue that this "inter-hemispheric imbalance" model arises from a fundamental misunderstanding of the roles played by inter-hemispheric (callosal) projections in mammalian brains. Drawing upon a large body of empirical data – derived largely from animal models, and associated theoretical modeling, it is demonstrated that inter-hemispheric projections perform contrast enhancing and integrative functions via mechanisms such as surround/lateral inhibition.  The principal functional unit of callosal influence comprises a facilitatory centre and a depressing peripheral zone, that together shape the influence of converging inputs to pyramidal neurons. Inter-hemispheric inhibition is an instance of a more general feature of mammalian neural systems, whereby inhibitory interneurons act not simply to prevent over-excitation but to sculpt the output of specific circuits. The narrowing of the excitatory focus that occurs through crossed surround inhibition is a highly conserved motif of transcallosal interactions in mammalian sensory and motor cortices. A case is presented that the notion of "inter-hemispheric imbalance" has been sustained, and clinical interventions derived from this model promoted, by erroneous assumptions concerning that revealed by investigative techniques such as transcranial magnetic stimulation (TMS). The alternative perspective promoted by the present analysis, also permits the basis of positive (e.g. post stroke) associations between the structural integrity of transcallosal projections and motor capability to be better understood.

Keywords: stroke; corpus callosum; cortex; brain stimulation; motor control

## Background

In seeking to deal cost-effectively with the burden of disability arising from stroke, national commissioning programs have emphasised a requirement for new rehabilitation technologies and novel modes of therapy (e.g. Department of Health, 2007). In this context non-invasive brain stimulation (NIBS) has come to prominence. Among the various modalities of NIBS, transcranial magnetic stimulation (TMS) and transcranial direct current stimulation (tDCS) have been investigated most extensively. It has been shown that in animal models, these forms of NIBS are capable of acting through a variety of cellular and molecular pathways that: change gene activity and protein expression; alter inflammatory processes and oxidative stress; influence cell migration and orientation; regulate neurotransmitter and neurotrophin release; and promote neurogenesis and angiogenesis (for reviews see Cirillo et al., 2017; Pelletier et al., 2015). When used in humans with the intent of ameliorating the impact of stroke, TMS and tDCS are typically applied with a view to altering cortical excitability - via the generation of action potentials or changes in resting membrane potential. This approach – in which the objective is typically to increase the excitability of neural circuits in the ipsilesional hemisphere, and/or decrease the excitability of circuits in the contralesional hemisphere, is predicated upon a specific conception of how the two sides of the brain interact. This has become known as the inter-hemispheric competition model.

At the core of this model is the assumption that during the generation of voluntary movement by one arm, there is inhibition of motor centres that give rise to efferent projections onto motoneurons innervating the homologous muscles of the opposite limb (e.g. Duque et al., 2005; Kobayashi et al., 2003; Netz et al., 1995). In referring to this assumption, the term "inter-hemispheric inhibition" is typically used. Although an interceding role of other pathways has not been precluded (e.g. Innocenti et al., 2017), the consensus view has always been that the fibres of the corpus callosum mediate inter-hemispheric inhibition (Meyer et al., 1995). In the context of the inter-hemispheric competition model, it is widely believed that damage to one cerebral hemisphere as a consequence of stroke leads to a reduction of inter-hemispheric inhibition from the damaged to the non-damaged hemisphere. As a corollary, the excitability of the non-lesioned hemisphere increases i.e. following release from inhibition by its counterpart. It is held that elevated excitability of the non-damaged side of the brain in turn brings about greater (inter-hemispheric) inhibition of the damaged side (Murase et al., 2004). Since it is assumed that this would have an adverse influence upon recovery of motor capability, the inter-hemispheric competition model has provided an impetus for the development of interventions for which the intent is to "rebalance" the hemispheres. As there exist specific forms of NIBS that have been shown to either increase or decrease the excitability of circuits within the primary motor cortex in healthy brains, it has been proposed that - if applied to reduce an imbalance of excitability between the hemispheres, they may provide a basis for stroke therapy (Boddington & Reynolds, 2017). In seeking to reappraise the function of inter-hemispheric inhibition, it is instructive to consider the level of success that has been achieved using interventions predicated on the conception of an "inter-hemispheric imbalance" – in particular those that have aimed to reduce inhibitory drive from the undamaged hemisphere to the damaged hemisphere.

**Effectiveness of interventions based on the Inter-Hemispheric Competition Model**

As there has been a steady accumulation of randomized controlled trials (RCTs) in which various forms of NIBS have been applied with the intent of decreasing the excitability of circuits in the contralesional primary motor cortex (M1), it has become feasible to undertake meta-analyses of the clinical outcomes. The forms of repetitive TMS (rTMS) examined most frequently in this context have been those in which either 1Hz trains (usually 1200 – 2400 stimuli in each application), or the "continuous" variant of theta burst stimulation (usually 600 stimuli in each application), have been applied. The meta-analyses reveal that, when compared to task-specific training alone, there is no additional benefit of 1 Hz rTMS applied over the contralesional M1 on upper limb impairment (Valkenborghs et al., 2019). A similar conclusion may be drawn on the basis of studies in which the lower limb has been the focus of attention (Tung et al., 2019). Even when considered relative to a sham stimulation control – rather than to usual therapy, there is little evidence of a positive effect of (inhibitory) continuous theta burst stimulation on measures of upper limb function (Zhang et al., 2017). Corresponding meta-analyses have been undertaken to quantify the effect of cathodal tDCS - delivered with the objective of decreasing the excitability of contralesional M1, prior to or during motor training.

These analyses have likewise failed to produce evidence that the addition of cathodal tDCS generates clinical outcomes superior to those achieved through rehabilitation training alone (Backhouse et al., 2018; Triccas et al., 2016). Indeed, even when compared only to sham stimulation, there is no demonstrable benefit of cathodal tDCS in terms of clinical measures of upper (Elsner et al., 2016, 2017) or lower extremity function (Elsner et al., 2016). It is also apparent that with respect to rTMS in particular, and non-invasive brain stimulation in general, the results included in most meta-analyses exhibit a prior "significance bias", such that the true effect sizes are likely to be very much smaller than those suggested by this method of aggregation (Amad et al., 2019). This inference is corroborated by the outcomes of an adequately powered multi-centre RCT ($\approx$ 200 participants) in which 15 minutes of 1 Hz rTMS (or sham rTMS) was delivered to the contralesional M1, prior to each of 18 sessions of rehabilitation therapy undertaken during an intervention period of six weeks. There were no differences between the rTMS and the sham groups with respect to any clinical measure of upper limb impairment or function, either immediately after the intervention period or during the six months following (Harvey et al., 2018). The more general conclusion to be drawn is that interventions delivered with the intent of inhibiting the contralesional M1, whether by means of rTMS or cathodal tDCS, do not have clinical utility in the context of movement rehabilitation following stroke (e.g. Nicolo et al., 2018).

As there is little evidence that decreasing the excitability of circuits in the contralesional M1 has a positive impact upon clinical outcomes, it is germane to consider the nature of the experimental studies that provided the foundation for this approach. In an influential investigation, Murase and colleagues (2004) reported that during the period immediately preceding the initiation of index finger movement, there was an inhibitory influence of the intact M1 on the lesioned M1 (in chronic survivors of subcortical stroke) that was in contrast to the facilitation seen in controls. It was also reported that the depth of the inhibition was correlated with assessments on the Medical Research Council (MRC) scale for muscle power. The key predictive measure was a paired pulse TMS elicited phenomenon that has also come to be known as "inter-hemispheric inhibition" (IHI). It refers to the decrease in the magnitude of a "test" motor evoked potential (MEP) that is obtained in response to stimulation delivered over M1, in circumstances in which an initial "conditioning" stimulus has been applied 6-15 ms previously to the opposite M1 (Ferbert et al., 1992). Hereafter in this paper, "inter-hemispheric inhibition" is used to refer to a physiological process, and "IHI" to refer to this electrophysiological phenomenon. There also exists a long latency (LIHI) variant that is elicited using inter-stimulus intervals (ISIs) in the region of 50 ms. Another such experimental index is provided by the ipsilateral silent period (iSP). This is elicited when TMS is delivered at high intensity to the M1 ipsilateral to contracting muscles (Wassermann et al., 1991). These measures have in common that the magnitude of the effect (i.e. the notional depth of inhibition) increases with the intensity of the conditioning stimulus, and with muscle contractions performed by the limb contralateral to the site of the CS (for a fixed CS intensity) (Perez & Cohen, 2008; Chiou et al., 2013). Numerous studies with similar objectives to those of the Murase et al. (2004) have been conducted including several that employed either the LIHI (e.g. Kirton et al., 2010) or the iSP measure (e.g. Takechi et al., 2014). Frequently the stated conclusions are aligned with the view that many stroke survivors exhibit elevated inter-hemispheric inhibitory drive from M1 in the non-damaged hemisphere to M1 in the damaged hemisphere (Ward and Cohen, 2004). In the following section, consideration is given to whether the clinical inferences made on the basis of such assertions are 1) conceptually valid, and 2) in accord with the accumulated empirical evidence.

**Reassessing the Inter-Hemispheric Competition Model**

It might first be determined whether there are logical deficiencies in the reasoning used to justify interventions undertaken with the aim of rebalancing excitability between the hemispheres, and by this means seeking to promote recovery of function. An important consideration is that: a change in IHI over time that correlates with clinical measures; variation in expression of IHI across individuals that correlates with variation in clinical presentation; or an intervention that gives rise both to a change in IHI and a change in clinical status, cannot be taken to indicate that there is a causal relationship between IHI and behavioural status (see also Bestmann & Krakauer, 2015; Xu et al., 2019).

A further critical issue is that IHI, LIHI and the iSP are electrophysiological phenomena that bear indirect relationships to multiple neurophysiological processes. Perhaps due to the (with hindsight unfortunate and ultimately confusing) adoption of the term "inter-hemispheric inhibition" (IHI) to denote a particular TMS technique, the discourse is frequently such as to suggest that IHI provides a direct index of a physiological process with which it shares the name. This is representative of a widespread, and often unwarranted, dependence on TMS derived measures in formulating more general statements concerning motor physiology (Carson et al., 2016). The reality is that the elicitation of all of these measures is such that they are relatively blind to, and indeed may mask, the expression of certain key physiological processes (Bestmann & Krakauer, 2015). This is a matter to which further consideration will be given in sections that follow.

Even if TMS derived measures are taken at face value, a damning indictment of the assumptions upon which interventions to rebalance the excitability of the cerebral hemispheres have been based is provided by the empirical data. In the context of a large-scale meta-analysis, McDonnell and Stinear (2017) reported that regardless of the TMS measure that was employed (IHI or iSP), there was no evidence of an imbalance between the hemispheres, when this was assessed in either acute or chronic stroke survivors. In other words, inhibition of the stroke-affected hemisphere by the non-affected hemisphere – as assayed by TMS, was not different from inhibition of the non-affected hemisphere by the affected hemisphere.

**Is there inhibition of homologous motor centres during voluntary movement?**

A vital point to be made in the present context is that during movements that are unilateral by intention, there is net facilitation – rather than net inhibition, of neural pathways projecting to homologous muscles of the opposite limb. If there is one task for which TMS is relatively well suited, it is in reflecting the incidence of variations in the state of short- and long-range projections onto large diameter corticospinal neurons with fast-conducting axons, which have cell bodies within M1, and innervate motoneurons in the spinal cord (Lemon et al., 2002). In circumstances in which either direct or indirect activation of such pyramidal tract neurons by TMS results in a descending corticospinal volley that gives rise to depolarization of spinal motoneurons, this can be registered as a MEP. Since unilateral contractions increase the amplitude of MEPs evoked in the opposite limb (Hess et al. 1986; Muellbacher et al., 2000; Stedman, Davey, & Ellaway, 1998), it can be inferred that unilateral actions induce bilateral elevations in corticospinal excitability (see Carson, 2005). This has been termed "crossed facilitation". If the evoked response is elicited by (electrical) stimulation at the level of the cervico-medullary junction (rather than by TMS over the primary motor cortex), corresponding changes in response amplitude are not necessarily obtained (Carson et al., 2004; Hortobágyi et al., 2003). This suggests that the changes in corticospinal excitability arise principally from alterations in the state of circuits within the cortex, rather than at the level of the spinal cord.

It can also be seen that the amplitudes of potentials evoked by TMS in quiescent muscles of a static limb increase with the level of tension generated by the corresponding muscle on the other side of the body (Bunday & Perez, 2012; Hortobágyi et al., 2003; Perez & Cohen, 2008, 2009). During oscillatory movements, the MEP is potentiated to the greatest degree during the phase of motion during which the homologous muscle of the opposite limb is maximally engaged (Carson et al., 1999; Carson et al., 2004; Stinear & Byblow, 2002). MEPs are potentiated in the period preceding the initiation of a movement by the opposite limb. This increase in the excitability of corticospinal projections from the ipsilateral (i.e. to the movement) M1 reflects the specific muscles that are engaged rather than the direction of the movement in space (Chye et al., 2018).

In short, the empirical evidence indicates emphatically that during movement of the upper limb, there is net facilitation, *rather than inhibition*, of motor centres that give rise to efferent projections onto motoneurons innervating the homologous muscles of the opposite limb. The most obvious conclusion to be drawn is therefore that a notional state of affairs whereby there is inhibition of one hemisphere by its counterpart does not prevail. A more general question is whether, from a phylogenetic perspective, the presence of such "crossed facilitation" is adaptive.

**What is the nature of the problem to which inter-hemispheric inhibition supposedly provides the solution?**

In seeking to account for the crossed facilitation that occurs during unilateral actions, it has been suggested that the phenomenon reflects a phylogenetically ancient organisation of the motor system that promotes symmetrical bilateral movements (e.g. Hopf, et al., 1974). If there is a predisposition of this nature, its expression does not appear to be mediated by direct anatomical connections between the cortical regions from which project the majority of fast-conducting corticospinal axons that innervate spinal motoneurons. Indeed, the sparse nature of the (direct callosal) projections between left and right caudal M1 in Old World monkeys and in humans (which may be designated posterior M1 (M1p) in histological atlases of the human brain) appears to represent one facet of a more recently evolved organization of motor cortex that supports fractionated movements of the digits and independent control of the hands (Rouiller et al., 1994; Wiesendanger et al., 1994). Since there is little scope (i.e. in terms of a material substrate) for direct facilitatory interactions between the left and right caudal portions of M1 (Dancause et al., 2015; Ruddy, Leemans, & Carson, 2017), by the same token, there is no obvious means by which direct inter-hemispheric inhibition between these specific areas can be instantiated. In other words, the term "last-stage inhibition" (Cincotta & Ziemann, 2008) does not apply. But what is the nature of the problem to which inter-hemispheric inhibition supposedly provides the solution?

In 1985, MacKenzie and Marteniuk noted that there is no complete taxonomy of bimanual coordination tasks (MacKenzie & Marteniuk, 1985). Notwithstanding classification of bimanual tasks for the purpose of defining post-stroke treatment targets (Kantak et al., 2017), more than a quarter of a century later this remains the case. There is however a degree of consensus in relation to their defining features. Cooperative goal directed actions performed by the two hands are considered emblematic of many uniquely human skills. Often in such tasks there is a clear differentiation of roles. Characteristically, the so-called "non-dominant" limb (i.e. the left in right handed individuals) performs a postural role, whereas the "dominant" limb acts upon or manipulates objects. In some cases the non-dominant hand provides a positional reference for the manipulative hand (Wiesendanger et al., 1994). The demarcations extend to motor skills for which many years of intensive practice are required to achieve mastery (Ericsson et al., 1993). For example, in playing the violin the dominant hand wields the bow; the non-dominant hand provides support and executes the fingering. When framed simply in terms of the goal of a task – such as strumming a chord on the guitar, the clear differentiation of roles accorded to the two hands at first glance implies that any intrinsic tendency towards symmetry of simultaneous movements must necessarily have been suppressed. What descriptions at this level of analysis fail to reveal is temporal segmentation. The picture that emerges from detailed observation is that the CNS formulates a single chain of timed commands in the context of which constituent movements are interleaved (Peters, 1977). Even for highly accomplished musicians, departure from a common time base (i.e. for the two hands) presents almost insurmountable difficulties (Peters, 1985). In short, for most cooperative goal directed actions in which the hands play differentiated roles, movements (transitions between postures) are sequential rather than simultaneous. When the two hands do perform movements *simultaneously*, the pervasive tendency is for these to be symmetrical.

This can be illustrated in many ways. In unimanual, target-directed aiming tasks, the size of the target, and the distance of the target from the starting position, predict movement duration in a fashion that is described well by Fitts' Law (Fitts & Peterson, 1964). If both hands are moved to separate targets for which Fitts' Law predicts distinct durations, the hands nonetheless start and end movement in close temporal proximity to one another (Kelso et al., 1979a, 1979b; Marteniuk et al., 1984; Riek et al., 2003). The disposition towards symmetrical simultaneous movement is also readily revealed by continuous/repetitive tasks, particularly when their rate of execution reduces the time available for the CNS to interleave limb specific sequences of motor commands. When asked to repeatedly draw circles with one hand and vertical lines with the other, spatial coupling occurs, whereby the trajectory of each limb adopts the features of the other (Franz et al., 1991). Patterns of bimanual coordination in which homologous muscles are engaged in an alternating fashion cannot be performed in a stable fashion, particularly as the frequency of movement is increased. In contrast, patterns in which homologous muscles are active simultaneously – and the movements therefore symmetrical, typically remain stable to the point at which the frequency of movement can no longer be increased (Cohen, 1971; Kelso, 1984; Riek et al., 1992). In the course of performing the tasks of daily living, such spatial and temporal coupling may be transient (Heuer, 1993), and often not readily perceived or observed. The tendency for simultaneous symmetrical movement is nonetheless ubiquitous (Swinnen, 2002). This is the context in which the functional role of inter-hemispheric inhibition must be assessed, rather than in some notional state in which synchronous bilateral activity is suppressed - a state that the CNS manifestly does not achieve.

**The concept of surround inhibition**

It is a central tenet of the inter-hemispheric competition model that movements of one arm are accompanied by the inhibition of motor centres with projections onto motoneurons innervating the homologous muscles of the opposite limb. It is clear however that this is not what occurs. In normal physiological conditions, there is in fact a high degree of bilateral coupling not only between brain centres linked most directly to efferent projections, but also within the somatosensory network (e.g. Dietz et al., 2015). There is thus a need to consider the functional role that is played by inter-hemispheric inhibition in the context of such coupling, rather than persisting with the view that it is somehow necessary in order to oppose the "default" state. From a broader phylogenetic perspective, it is difficult to discern circumstances in which a system would have evolved to modulate excitation in such a manner that a *counteracting* layer of inhibition must then be superimposed. The alternative considered here is that inter-hemispheric inhibition is an instance of a more general feature of mammalian neural systems, whereby inhibitory interneurons act not simply to prevent over-excitation but to sculpt the overall output of specific circuits (Merchant & Georgopoulos, 2017).

Experiments in cat reveal the presence of facilitatory connections with the homotopic area of the opposite motor cortex, which are surrounded by a more extensive zone in which inhibitory responses to transcallosal stimulation are obtained (Asanuma and Okuda, 1962; see Figure 1). The wider dispersion of inter-spike intervals – registered by single unit recordings, suggests that there are a greater number of interceding synaptic relays in the peripheral (inhibitory) zone (Kogan and Kuraev, 1976). This accords with the consensus view that callosal neurons are typically glutamatergic (Innocenti, 1986; Werhahn et al., 1999) and exert a facilitatory influence upon their immediate targets in the opposite hemisphere (Conti & Manzoni, 1994; Houzel and Milleret, 1999; Voigt et al., 1988). Although callosal fibers arising from GABAergic cells have in some case been identified (e.g. Gonchar et al., 1995; Fabri and Manzoni 2004; Rock et al., 2018), for the most part the inhibitory responses arise from neural interactions that occur within circuits local to the opposite hemisphere (Bianki and Shrammapril, 1985; Daskalakis et al., 2002; Carson, 2005; Berlucchi, 1990). Only a small fraction of callosal inputs onto pyramidal cells are however subject to GABAergic (i.e. inhibitory) modulation (Carr & Sesack, 1998). This is consistent with indications that between 5% and 20% of long-distance extrinsic connections within cortex synapse onto inhibitory cells (Fisken et al., 1975; McGuire et al., 1991). More pertinently one might ask: what purpose do local inhibitory effects brought about by callosal input serve?

_______________________________________________

Insert Figure 1 about here

_______________________________________________

The organization described by Asanuma and Okuda bears some of the hallmarks of lateral/surround inhibition (Hartline & Ratliff, 1972) – a fundamental characteristic of neuronal processing. Barlow (1953) described surround inhibition in studies of ganglion cells in the frog retina. Barlow noted that the inhibitory action of light falling outside the receptive field of receptor cells exerts a discriminatory action on the response profile of the ganglion cells to which they project, which serves to increase the degree of contrast sensitivity that can be achieved. As Barlow concludes (p. 87), a system configured along these lines acts as a filter, "rejecting unwanted information and passing useful information".

It was established subsequently that this form of organization is omnipresent in sensory processing systems. For example, within the posterior column medial lemniscal system (PCMLS), surround inhibition is operative initially within the posterior column nuclei, and remains evident in all relays of the PCMLS. By this means the discrimination of stimuli applied to separate points on the skin is sharpened and enhanced – a facility necessary for capabilities such as two-point discrimination. Surround inhibition is also thought to play a significant role in sharpening the orientation selectivity of cells in primary visual cortex (V1) (Shapley et al., 2003). The columnar arrangement found in the cerebral cortex assumes particular functional significance in this regard (Mountcastle, 1957), supporting orderly patterns of connectivity that, through surround inhibition, not only focus but also amplify sensory input (Eccles, 1971).

Although historically the thesis has been subject to less extensive empirical scrutiny (e.g. Stefanis & Jasper, 1964a; 1964b), Eccles (1971) pointed out that the amplification and sharpening of neuronal activity realised by surround inhibition also assumes a key role in shaping motor output. This idea has been developed recently by Georgopoulos and colleagues, in a series of papers (Georgopoulos & Stefanis, 2010; Merchant et al., 2008) outlining the proposition that these local inhibitory mechanisms are critical for controlling the directional accuracy and speed of reaching movements (Mahan & Georgopoulos, 2013) (Figure 2). The observation that the local circuit mechanisms that influence the coding of movement in the motor cortex are in some cases similar in character to Renshaw inhibition (for review see Hultborn et al., 1979) in the spinal cord (Georgopoulos & Carpenter, 2015; Kameda et al., 1969; Stefanis & Jasper, 1964b) serves to emphasize that, as with its sensory counterpart, surround inhibition is a ubiquitous feature of the motor system.

_______________________________________________

Insert Figure 2 about here

_______________________________________________

Following Bianki (e.g. Bianki, 1981; Bianki and Makarova, 1980; Bianki and Shrammapril, 1985), we have proposed previously (Ruddy & Carson, 2013) that the organisation first described by Asanuma and Okuda (1962) - excitation by stimulation of an approximately symmetrical point in the opposite M1 cortex and inhibition by stimulation of the immediately surrounding area, provides for the sharpening of movement related neuronal activity through *crossed* surround inhibition. Although callosal fibers may innervate both pyramidal cells and inhibitory interneurons (e.g. Carr & Sesack, 1998; Karayannis et al., 2007), it is significant that the (GABAergic) cells mediating transcallosal (usually disynaptic) inhibitory post synaptic potentials (IPSPs) can be driven by axon collaterals of pyramidal cells that are excited monosynaptically by callosal projections (Conti & Manzoni, 1994). This is also a feature characteristic of lateral/surround inhibition. As will become apparent however, "crossed surround inhibition" is not a unique or anomalous phenomenon. Rather, it is fundamental to bilateral cortical interactions (Tang et al., 2007). It is only when viewed in this light that the adaptive utility of inter-hemispheric inhibition is revealed (Cook, 1984).

**Crossed Surround Inhibition: Sensory**

Jones and Powell (1969) proposed that callosally mediated surround-inhibition assumes a role in sensory discrimination that is equivalent in importance to that assumed by ascending sensory pathways. Although relative "importance" is a challenging attribute to gauge, it is nonetheless clear that crossed surround inhibition arising from callosal projections between homotopic regions is a feature of many sensory processing regions in primate and non-primate mammalian brains. A pattern of callosal effect giving rise to a facilitatory centre and a depressing periphery is evident in the visual and auditory cortices of cat (Bianki, 1981). Surround mechanisms mediated by transcallosal connections have also been demonstrated for several vision processing areas in various primate species (Allman et al., 1985). In model systems, it has been shown that they accentuate orientation tuning, directional selectivity, and velocity sensitivity (Simmons & Pearlman, 1983). Beyond the widely appreciated contribution to the generation of receptive fields close to the vertical midline – through point-to-point retinotopic correspondences across the hemispheres (e.g. Rochefort et al., 2009; Schmidt, 2016), callosal projections in visual cortex also appear to play further integrative roles. There are indications that neurons in primary visual areas with bi-hemispheric collaterals provide, through temporal coupling of focal excitatory activity induced in their (bilateral) post-synaptic targets, for the grouping of object features on the basis of temporal correlation (Houzel et al., 2002).

As for other cortical regions, callosal axons largely have an excitatory influence upon their immediate targets in the opposite somatosensory cortex (Sloper and Powell, 1979), some of which are inhibitory interneurons (Somogyi et al., 1983). The surround inhibition thus realised contributes to the shaping of neuronal receptive fields (RFs) (Iwamura et al., 2001; Clarey et al., 1996). It also plays an integrative role, promoting bilateral interactions in contexts in which these are functionally relevant – such as in determining the orientation of an obstacle (Shuler et al., 2001).

The auditory cortex is of particular interest. Whereas for cortical regions that mediate vision (Newsome & Allman, 1980) and somatosensation (Iwamura et al., 2001) callosal fibres largely (but not exclusively) innervate the midline zone, inter-hemispheric projections are distributed across the full extent of the auditory cortex (Bamiou et al., 2007; Kitzes & Doherty, 1994). In connecting homotopic and heterotopic cortical areas, it is thus redolent of the organisation that characterises the cortical motor network (Ruddy et al., 2017). There exist at least two classes of microcircuit in layer 5 of mouse auditory cortex that mediate responses to callosal inputs. Callosal projections onto corticocortical pyramidal neurons induce direct excitation and feedforward inhibition, whereas inputs to corticocollicular pyramidal neurons result in direct excitation (Rock & junior Apicella. 2015). A particular role in extracting and refining the complex temporal information derived from natural sounds has been ascribed to the coordinated processing by both hemispheres made possible by the feedforward inhibition to which these projections give rise (e.g. Rock & junior Apicella, 2015; Villa et al., 2007). As in other areas of cortex, recurrent collateral axons are also instrumental in mediating the inhibition observed in the immediate vicinity of neurons that receive callosal projections from the homotopic regions of the contralateral hemisphere (Kitzes & Doherty, 1994). A further integrative function has been ascribed to crossed surround inhibition identified in the auditory cortex (e.g. Blackwell & Geffen, 2017; Irvine et al., 1996). In bats for example, neurons in this region are responsive to the combination of the emitted biosonar pulse and its echo - with a specific echo delay. Balancing the delay maps of the two hemispheres in the analysis of orientating echo sounds is accomplished via bilateral interactions that comprise focal facilitation and widespread lateral inhibition (Tang et al., 2007).

**Crossed Surround Inhibition: Motor**

If, as has been suggested (e.g. Tang et al., 2007), bilateral interactions of this nature are among the most fundamental mechanisms of cortical processing, they should also be found in motor systems. In this regard, the findings of Asanuma and Okuda (1962) have been extended on the basis of studies in a number of mammalian species. For example, Chapman et al. (1998) reported that in rat, the initial monosynaptic response to callosal input in layer V neurons of M1, is followed by polysynaptic activation of a much larger and more spatially diffuse population of neurons. The inhibitory action of these cells, for example those located within layer II–III, is believed to contribute to the potentiation of a later component of the response generated by monosynaptic input to layer 5, in a fashion that serves to narrow the extent of motor output. The organization described by Chapman et al. is thus consistent with the more general conjecture that transcallosal projections onto pyramidal neurons in the deep cortical layers of M1 serve to enhance the excitatory focus of other (e.g. thalmocortical) inputs (Shramm & Kharitonov, 1984).

The explanatory power of the studies conducted by Asanuma and Okuda (1962) is derived from the carefully titrated stimulation of pyramidal tract cells in homotopic regions of the opposite cortex, rather than (as is otherwise frequently the case) direct stimulation of callosal fibres. As Asanuma and Okuda (1962) point out, if a relatively large number of callosal fibres are activated (by whatever means), the focal excitatory effect is masked by the overlapping and summative influence of those fibres which project onto interneurons in the opposite hemisphere that mediate the expression of surround inhibition. Although they cannot provide the spatial or temporal specificity of the methods employed by Asanuma and Okuda, complementary information can be derived from studies in which intracortical microstimulation (ICMS) is used to assess changes in motor output following (e.g. lidocaine-induced) suppression of the homotopic area in the opposite hemisphere. In this context, the vibrissa motor cortex (VMC) in rats is employed frequently as a model system. On the basis of a series of studies of this nature (e.g. Maggiolini et al., 2007; 2008), it was surmised that stimulated sites in receipt of facilitatory input via the callosum assume a particular role in preserving the size, shape and excitability of vibrissal representations. For those sites subject to an inhibitory influence of the opposite hemisphere, a role in maintaining a boundary between representations of the vibrissa and the forelimbs was reported. The view that the balance between crossed (focal) facilitation and (lateral/surround) inhibition helps maintain the definition of representations within motor cortex is further supported by studies in which preservation of the corpus callosum prevents encroachment by representations of the vibrissae and the hindlimb into areas with projections to muscles innervated via the brachial plexus, that is otherwise observed following total brachial plexus root avulsion (Zhang et al., 2015).

**Functions and features of Crossed Surround Inhibition**

The contrast enhancing function of surround/lateral inhibition is a motif of transcallosal interactions in the sensory and motor cortices. The principal functional unit of callosal influences - comprising a "facilitatory center and depressing periphery", serves as a "sculptor's chisel" (attributed by Bianki, 1981 to Ukhtomskii, 1966), that shapes the influence of converging inputs to pyramidal neurons. Critically, the narrowing of the excitatory focus that occurs through this means is reciprocal in nature (Bianki and Makarova, 1980). Increases in surround inhibition in one hemisphere give rise to a reverse (i.e. symmetrical and selective) influence on neurons in the contralateral hemisphere (Bianki and Shrammapril, 1985). The reciprocal nature of callosal connections is also central to arguments that they perpetuate *intrahemispheric* function - extending onto the contralateral hemisphere intrinsic networks of intracortical short- and/or long-range connections (Schmidt, 2016; see also Bianki, 1981). Thus conceived, the recruitment of contralateral ensembles increases the computational resources available to perform a given task (Ferrandez, 2016). These functions are not however exclusive. An architecture that, for example, provides for the sharpening of movement related neuronal activity through crossed surround inhibition necessarily also promotes an "integrative" propensity for simultaneous symmetrical movement. Modeling key features of the supporting neural architecture can also delineate the system properties that emerge as a consequence of crossed surround inhibition.

---

Insert Figures 3 & 4 about here

---

O'Hashi et al (2018) developed a neural mass model of the columnar organization of early visual cortex in which each cortical site is characterised as a local pool of inhibitory interneurons connected mutually to an excitatory pool of (pyramidal) neurons. Cortical sites were connected laterally (both within and across hemispheres) via excitatory projections.  Subject to reasonable assumptions that callosal connections are: relatively fast (Aboitiz, 2017; Innocenti et al., 1995); topographically precise (Innocenti, 1986; Rochefort et al., 2007); and have synaptic efficiency roughly equivalent to intrahemispheric projections, a number of attributes were exhibited by the model.  Spontaneous (emerging as under anesthesia) activity patterns corresponding to maps associated with cardinal and oblique orientations registered experimentally, and orientation state transitions were obtained. In this context, the functional contributions of cortical sites in the two hemispheres were such as to suggest that they were coupled as a single unit. The observation that inter-hemisphere synchrony was equivalent to intra-hemisphere synchrony was taken to further imply that the inter-hemispheric connections are as functionally efficient as intrahemispheric connections (cf. Jones and Powell, 1969). Notably, the maps corresponding to cardinal and oblique orientations emerged as transient attractor states exhibiting features similar to those ascribed to the pattern dynamics of bimanual coordination. By virtue of extending the computational resources that are available, the distribution of task related neural activity across the two hemispheres also increases robustness. During motor planning, coupled (via callosal projections) preparatory activity that predicts specific movement directions can be registered in premotor cortex neurons in both hemispheres. If premotor circuits in only one hemisphere are silenced (using optogenetic methods), recovery of task specific neural activity occurs rapidly – driven by input from the opposite hemisphere. In contrast, bilateral photoinhibition disrupts preparatory activity. In computational models of this preparation, it is the presence of reciprocal callosal connections in particular that is required to reproduce the stability observed experimentally (Li et al., 2016).

The traditional inter-hemispheric competition model relies on the conception that GABAergic interneurons subject local pyramidal neurons to inhibition that is promiscuous and undifferentiated (e.g. Fino and Yuste, 2011; Packer and Yuste, 2011). With respect to the inhibition that is mediated by callosal fibres however, it is apparent that this is in fact highly differentiated, and reflects the functional specialisation of neurons (Lee et al., 2014) and local circuits in the various cortical regions to which these fibres project (Ferrandez, 2016).

Several forms of differentiation can be discerned. At the most embracing level, each area of cortex projects callosal fibres not only to homotopic regions in the opposite hemisphere, but also to heterotopic regions (Chovsepian et al., 2017; Innocenti, 1986; Miller & Vogt, 1984; Ruddy et al., 2017). The projection patterns are in most cases reciprocal. Although homotopic and heterotopic projections can arise from distinct cortical layers, in some cases individual neurons bifurcate axons to more than one region of the opposite hemisphere (Innocenti, 1986). With respect to the different laminae in which the projections terminate, the alterations in the balance between excitation and inhibition that are brought about by callosal input can vary markedly. Furthermore, the differences between layers in this respect do not remain constant across cortical regions. For example, in rat somatosensory cortex, $GABA_B$ receptor mediated inhibition of layer 5 pyramidal neurons, via transcallosal projections, appears to act upon on the apical dendrites via layer 1 interneurons. In this preparation, layer 2/3 pyramidal neurons are not subject to inhibition by callosal input (Palmer et al., 2012; 2013). In contrast, recordings from retrosplenial (RSC) cortex in mice suggest that layer 2/3 pyramidal neurons are inhibited profoundly by callosal input acting via parvalbumin (PV) expressing gabaergic interneurons, whereas thick-tufted but not thin-tufted pyramidal neurons in layer 5 were subject to excitation (Sempere-Ferrández et al., 2018).

The distinct response profiles registered for thick-tufted and thin-tufted pyramidal neurons in layer 5 points to further basis for differentiation. Specifically, distinct classes of projection neurons exhibit different changes in the balance between excitation and inhibition brought about by callosal input. The levels of callosally evoked excitation and feedforward inhibition that can be recorded in layer 5 of mouse prelimbic prefrontal cortex (PFC) are greater in cortico-thalamic (CT) neurons than in cortico-cortical (CC) neurons. Although in both cases the inhibition is brought about largely by PV expressing GABAergic interneurons that receive monosynaptic, glutamatergic input from the contralateral hemisphere, the differentiated response profiles are determined, at least in part, by differences in the intrinsic physiology of the CC and CT neurons (Anastasiades et al., 2018; see also Lee et al., 2014). By way of contrast, in (mice) auditory cortex, a similar pattern of variation between the response profiles of layer 5 corticocortical (CCort) and corticocollicular (CCol) neurons, appears to be due to a preferential (feedforward) activation of CCort pyramidal neurons by PV expressing inhibitory interneurons. In the absence of strong suppression by inhibitory interneurons, the CCol projecting neurons exhibit greater facilitation in response to callosal input (Rock & junior Apicella, 2015). In this case therefore, the variations in the response profiles of the two types of pyramidal neuron are brought about by differences in the architecture of the networks in which they are imbedded, rather than being attributable to a distinction in their intrinsic physiology.

In mouse the RSC forms part of a subnetwork that mediates the transduction of information between sensory areas and higher-order association areas (Zingg et al., 2014). Callosal fibres excite monosynaptically layer 6 pyramidal neurons in the agranular subdivision of the RSC. In addition, PV expressing GABAergic interneurons, in this deep cortical layer, receive synapses on dendrites and somata and are activated directly by the contralateral hemisphere via callosal projections (Karayannis et al., 2007). The direct (feedforward) callosal activation of GABAergic interneurons observed in this case (similar to that shown in Figure 3) can be differentiated from an organisation whereby inhibitory postsynaptic potentials (IPSPs) are driven by axon collaterals (Figure 4) of the pyramidal cells excited monosynaptically by callosal projections (Conti & Manzoni, 1994). In the period since the first demonstrations of recurrent collateral inhibition in pyramidal tract neurons (e.g. Phillips, 1959; Stefanis & Jasper, 1964; Kameda et al., 1969), it has been proposed that these recurrent collaterals, and the intercalated interneurons onto which they project, provide the substrate for the bell-shaped directional tuning curve that characterises the discharge of pyramidal cells in motor cortex (Georgopoulos & Stefanis, 2007). Indeed, a substantial proportion of individual (presumed GABAergic inhibitory) interneurons within primate M1 can be classified on the basis of their directional tuning (Merchant et al., 2008). In principle, while reductions of discharge time and constriction of excitatory extent induced by collateral inhibition provide for both temporal and spatial enhancement of motor output (Isomura et al., 2009; Kameda et al., 1969), it has also been argued that feedback inhibition and feedforward inhibition assume complementary computational roles (Chena et al., 1995). In this scheme, feedforward inhibition sharpens the focus of motor cortical activation to a greater degree, whereas feedback inhibition increases the temporal precision of motor output. With respect to callosal input more particularly, the temporal dynamics of the inhibition that it brings about are also determined by receptor subtype. In the context of direct (feedforward) inhibition, the IPSPs generated in inhibitory interneurons following transcallosal monosynaptic EPSPs have a component that depends on $GABA_A$ receptors, and another that is contingent upon $GABA_B$ receptors (Conti & Manzoni, 1994; see also Chowdhury & Matsunami, 2002).

In summary, and consistent with the general observation that callosal projection neurons are extraordinarily diverse (Fame et al., 2011), the local inhibition that these neurons may promote is differentiated in many ways. This variegation arises not only from cytoarchitectural features, such as the brain regions and cortical layers in which the cell bodies of callosal projection neurons are located, it is also attributable to the intrinsic morphology and physiology of the target cells onto which they project, and the configuration of the networks in which these targets are imbedded. In short, the recruitment profile of inhibitory interneurons engaged (directly or indirectly) by callosal input is extremely complex, and belies any model in which the function of inter-hemispheric inhibition is represented as simply modulating the overall excitability of brain networks. As Cook (1984) perceived more than thirty-five years ago, the existence of inter-hemispheric inhibition does not suggest that unilateral activity generates contralateral "deactivation", or that it brings about asymmetries of excitation between the cerebral hemispheres. Rather, acting through the mechanism of crossed surround inhibition, it provides the basis for "*complementary* patterns of neuronal firing" (page 121).

**Limitations of the techniques used to investigate Inter-hemispheric Inhibition in humans**

In a fitting observation, Uttal (2016) notes that, "our theories are, to a much greater degree than we appreciate, creatures of whatever technology is available at any stage of history" (page xii). In this vein, it will be argued that the inter-hemispheric competition model has been sustained, and clinical interventions based on the notion of an "interhemispheric imbalance" promoted, by a misplaced dependence upon experimental methods that depend upon TMS. More specifically, erroneous assumptions have been made concerning that which can be revealed by the IHI technique.

As emphasised previously, IHI is an electrophysiological phenomenon whereby the magnitude of a "test" motor evoked potential (MEP) - obtained in response to stimulation delivered over M1, is decreased by the application of a prior "conditioning" stimulus (CS) to the opposite M1 (Ferbert et al., 1992). The intensity of the CS employed to elicit IHI is in almost all instances well above that required to generate a MEP in quiescent muscles of the opposite limb. For example, Ferbert et al. (1992) used a CS that was 20% above the motor threshold. When varied systematically in relation to the active motor threshold (AMT), it becomes apparent that (short and long latency) IHI can be invoked only when the CS intensity is well above (e.g. > 1.2 AMT) the level at which descending corticospinal volleys are generated (see Figure 6 in Ni et al., 2008).

A principal limitation arising from the use of TMS in this context is that it has extremely poor spatial definition. When modeled on the basis of MEPs generated by direct electrical stimulation (DES) of the exposed cortex in individuals undergoing neurosurgery, TMS delivered at 120% of resting motor threshold (RMT) exerts a stimulating effect that can extend over one or two neighboring gyri, and span an area of several $cm^2$ (Opitz et al., 2014). To put this in context, estimates of the number of neurons within a unit volume below 1 $mm^2$ at the surface of the primary motor cortex in chimpanzee are in the region of 55,000 (Young et al., 2013). It is clear therefore that millions of neurons are likely to be affected directly by TMS. Since cortico-cortical callosal projections terminate both in homotopic subregions of the opposite hemisphere (Innocenti et al., 2017), and also in heterotopic regions (Innocenti, 1986; Chovsepian et al., 2017), there is no reason to suppose that in the IHI paradigm, the transcallosal effect of the CS will be any more closely circumscribed. Indeed, it has long been appreciated that even when DES is applied (for example Opitz et al., 2014 induced a stimulation area of a few mm$^2$ using currents below 20 mA in humans), the spatial extent of the callosal stimulating effect is very much greater than under natural conditions (Doty & Negrao, 1973). It is also clear that both magnetic and electrical stimulation of M1 induce high-frequency (~600 Hz) repetitive discharge in pyramidal neurons, which is quite distinct from normal physiological discharge frequencies of 100 impulses/s (Lemon & Kraskov, 2019). As a consequence, even a single discrete (artificial) stimulus can invoke trans-synaptic responses in the majority of neighbouring neurons (Maier et al., 2013). A key observation in this regard is that, while relatively weak electrical stimuli have an excitatory effect, as the effective intensity of the stimulation is increased, inhibition that affects "an astonishingly large number of neurons" can be generated (Krnjević et al., 1966). This is a characteristic of both intracortical and transcallosal stimulation (Asanuma & Okamoto, 1959; Krnjević et al., 1966).

The practical consequences of these factors in relation to IHI have been recognised for some time (Bäumer et al., 2006; Hanajima et al., 2001). In short, the poor spatial definition of the transcallosal volley brought about by the conditioning magnetic stimulus, and the intense transsynaptic bombardment of neurons to which TMS gives rise, both promote the massed and undifferentiated invocation of inhibitory processes that in normal physiological conditions are precisely engaged and serve to narrow the excitatory focus of M1 activation. Paradoxically therefore, in applying the IHI technique, the defining and integrating inter-hemispheric interactions that occur through crossed surround inhibition, are masked by the artificial means through which the supporting neural networks are interrogated. As a consequence of the disproportionate significance ascribed to results obtained using the IHI method, the adaptive functions of inter-hemispheric inhibition may have been largely misconceived.

**Corroborating clinical evidence**

It is a central tenet of the inter-hemispheric competition model that the contralesional hemisphere brings about inhibition of the ipsilesional hemisphere via callosal pathways that link the two sides of the brain. In this scheme, it follows that callosal integrity (i.e. reflecting the quality of the medium through which inter-hemispheric interactions occur) should be positively related to the extent of inhibition, and thus to the magnitude of the motor deficits that are exhibited following stroke.  In other words, better callosal connectivity should, in the terms of this model, provide greater scope for inhibition of the ipsilesional hemisphere by the contralesional hemisphere. If in contrast, inter-hemispheric projections serve primarily to focus and integrate neural activity in the opposite side of the brain via mechanisms such as surround inhibition, callosal integrity in stroke survivors should be positively related to motor capability. That is, greater, or improved, callosal connectivity should entail better motor capability. There is a steadily accumulating body of empirical evidence – derived largely from neuroimaging studies, that provides a basis upon which to assess these contrasting predictions. Due to the diversity of the imaging methodologies and analysis techniques that have been applied, meta-analyses are not yet available. It is however possible to provide a narrative summary of the key findings. These are consistent in indicating that the structural integrity of transcallosal pathways is positively associated with motor capability following stroke.

The MRI variant referred to as diffusion weighted imaging (DWI) is capable of providing quantifiable detail concerning the microstructural organisation of white matter, and in some implementations it permits the course of individual white matter tracts to be resolved. The diffusion tensor model first described by Basser et al. (1994), is the mathematical framework that has been used most widely to make inferences that relate diffusion-weighted images to the local tissue microstructure. The model is however inadequate for regions that contain complex architectures such as crossing fibres (Alexander et al. 2002; Jeurissen et al. 2013; Tuch et al. 2002) – a limitation that extends to the lateral cortical projections of the corpus callosum (Jeurissen et al. 2011). As many of the transcallosal projections that connect nodes of the cortical motor network cannot therefore be detected using the diffusion tensor model (Meng & Zhang, 2014; Ruddy et al., 2017), the present summary is restricted to studies that have used non-tensor diffusion models, such as constrained spherical deconvolution (CSD) imaging and diffusion spectrum imaging (DSI).

Assessing the fractional anisotropy (FA) – higher values of which are obtained for brain regions that are heavily myelinated or that have densely packed axons, of tracts defined within the midsagittal corpus callosum using the CSD method, Auriat et al. (2015) reported a negative association between FA values and completion times for the Wolf motor function test (WMFT), in twenty-seven chronic stroke survivors who ranged widely in the degree of arm impairment that was exhibited (Fugl-Meyer upper limb score (FM-UL) 7 to 63). In a more severely impaired (FM-UL ≤ 30/66) group, Hayward et al. (2017) observed that CSD derived FA values for callosal streamlines defined between 1) prefrontal; 2) premotor; and 3) primary motor regions were in all cases positively associated with FM-UL score. It does however appear that for chronic stroke survivors, the negative associations of FA values with motor function (e.g. WMFT completion time), and the positive associations with the FM-UL score, are more robust for callosal streamlines linking prefrontal regions and premotor/SMA regions, than those defined for primary motor regions (Hayward et al., 2017; Mang et al., 2015). The FA values derived for callosal fibres projecting between primary sensory, and parietal, temporal and visual areas, seem to have minimal predictive power in relation to measures of either motor function or impairment (Mang et al., 2015). Using DSI tractography, Koh et al. (2018) observed that in a group of chronic stroke survivors with lesions impinging on callosal motor fibres, there was a positive association between FM-UL score and the generalized fractional anisotropy (GFA) measure derived for streamlines passing between homologous regions of the cortical motor network.

Although changes in the diffusion characteristics of white matter bundles are substantially less marked than those observed for cortical grey matter in the days immediately following middle cerebral artery (MCA) occlusion (Zhang et al., 2018), decreases in the microstructural integrity of callosal fibres distal to the focal lesion (i.e. Wallerian degeneration) are evident within six months following MCA and (subcortical) pyramidal tract infarcts (Gupta et al., 2006; Radlinska et al., 2012; see also Li Y. et al., 2015, 2016). Measurements of structural connectivity obtained in the sub-acute phase following stroke, therefore provide the best possible basis upon which to resolve the presence of associations between the structural integrity of transcallosal pathways and motor capability following stroke. This is particularly the case if the observations are made within the time frame during which Wallerian degeneration of callosal fibres is unlikely to have occurred to a significant degree. It might be remarked that assessments of variations in functional connectivity (for example derived using resting state fMRI) are of limited utility in this regard, as these measures are necessarily sensitive to the immediate changes in the dynamic properties of neural circuits within the ipsilesional hemisphere that occur as a direct consequence of stroke. It is however feasible to evaluate whether, during the sub-acute phase, motor capability is positively related to the structural connectivity of transcallosal projections.

In a group of patients for whom the DWI was typically undertaken within the first ten days following stroke, probabilistic tractography derived indices of the white matter tract volumes defined between regions of interest (ROIs) encompassing left and right M1, were predictive of motor outcomes (box and block test) assessed three and six months later (Lindow et al., 2016). Employing DSI tractography, Granziera et al. (2012) noted that GFA values of white matter streamlines passing through left and right SMA, registered within one week following stroke, predicted scores on the motor part of the NIH Stroke Scale (NIHSS) obtained six months later. Using DWI sequences with an angular resolution lower than that required for the use of methods such as CSD, Li and colleagues performed probabilistic tracking of streamlines seeded in left and right M1, and reported a positive association between FA values and Fugl-Meyer scores for individuals having sustained subcortical ischemic lesions 21 to 135 (median = 32) days previously (Li Y. et al., 2016, see also Li Y. et al., 2015).

In short, in stroke survivors, the structural integrity of callosal fibre bundles that connect nodes of the cortical motor network is positively related to motor capability. The association is expressed both when assessed soon (< 10 days) after the brain insult, and many months later. This pattern of outcomes is consistent with the proposition that the callosal projection neurons contained within these fibre bundles support the focusing and integration of neural activity in the opposite side of the brain.

**Summary and Conclusions**

The practical and explanatory limitations of the "inter-hemispheric imbalance" model – as it has been applied to stroke, arise from a fundamental misunderstanding of the functions assumed by inter-hemispheric (callosal) projections in mammalian brains. These projections do not give rise to promiscuous and undifferentiated inhibition, or bring about asymmetries in excitation between the cerebral hemispheres. Rather, they perform contrast enhancing and integrative roles via mechanisms such as surround/lateral inhibition, that shape the influence of converging inputs to pyramidal neurons. The narrowing of excitatory focus promoted by inter-hemispheric inhibition is a functional motif that is strongly conserved across mammalian sensory and motor cortices. It is a highly differentiated process that reflects the functional specialisation of neurons and local circuits in the various cortical regions to which callosal fibres project. Any attempts to manipulate inter-hemispheric inhibition for therapeutic gain must – in order to be successful, respect its cardinal properties.

**Acknowledgements**

The author thanks Apostolos Georgopoulos and Winston Byblow for their generous and constructive commentary.

**Funding**

No specific funding was allocated for this work.

**Competing Interests**

There are no competing interests of which the author is aware.

**Figure Legends**

1) Author's rendition of Asanuma and Okuda (1962, Figure 9), representing the manner in which callosal projections from one hemisphere give rise to an excitatory influence upon pyramidal tract cells in a restricted area, and an inhibitory influence upon pyramidal tract cells in the surrounding region of the opposite hemisphere. Cells that generate excitatory postsynaptic potentials (EPSPs) are shown in black. The red cell is an inhibitory interneuron. The pyramidal cell labelled P is subject to excitatory drive. The pyramidal cell labelled $P^i$ lies within the peripheral inhibitory zone. Asanuma and Okuda assumed that a "re-exciting interneuronal chain" permits the interneurons to fire repetitively.

2) Author's representation of the hypothesis that the directional accuracy and speed of upper limb movements are achieved via circuits tuned variably by local inhibitory mechanisms, as described and illustrated in Georgopoulos and Carpenter (2015) and Mahan and Georgopoulos (2013). Afferent inputs (excitatory) are shown in cyan. The red cells are inhibitory interneurons. The pyramidal cell labelled P is subject to focused excitatory drive. The other pyramidal cells make up the "inhibitory fringe". It is hypothesised that "strong input" to inhibitory interneurons (A) sharpens the locus of motoneuron excitation by attenuating the contribution of the excitatory fringe. This results in a reduction of the directional tuning width and the generation of an accurate and short population vector. In contrast, "weak input" to inhibitory interneurons (B) results in an increase in the directional tuning width and the generation of a less accurate and longer population vector – corresponding to a faster but less accurate movement.

3) A schematic representation of the manner in which the contrast enhancing and integrative functions of crossed surround inhibition may be instantiated via direct callosal activation of GABAergic interneurons. Pyramidal tract cells are shown in black. The pyramidal cells labelled P are subject to focused excitatory drive. The red cells are inhibitory interneurons. Callosal projections (excitatory) are shown in cyan. The inhibitory influence upon pyramidal tract cells in the surround region of the opposite hemisphere ("inhibitory fringe") in this case arises from local *feed-forward* inhibitory circuits. A key assumption is that the inter-hemispheric interactions arising from this configuration can be reciprocal (e.g. Bianki and Makarova, 1980; O'Hashi et al., 2018).

4) A schematic representation of the manner in which the contrast enhancing and integrative functions of crossed surround inhibition may be instantiated via axon collaterals of the pyramidal cells excited monosynaptically by callosal projections. Pyramidal tract cells are shown in black. The pyramidal cells labelled P are subject to focused excitatory drive. The red cells are inhibitory interneurons. Callosal projections (excitatory) are shown in cyan. The inhibitory influence upon pyramidal tract cells in the surround region of the opposite hemisphere ("inhibitory fringe") in this case arises from local *recurrent* inhibitory circuits. A key assumption is that the inter-hemispheric interactions arising from this configuration can be reciprocal (e.g. Bianki and Makarova, 1980; O'Hashi et al., 2018).

# Figure 1

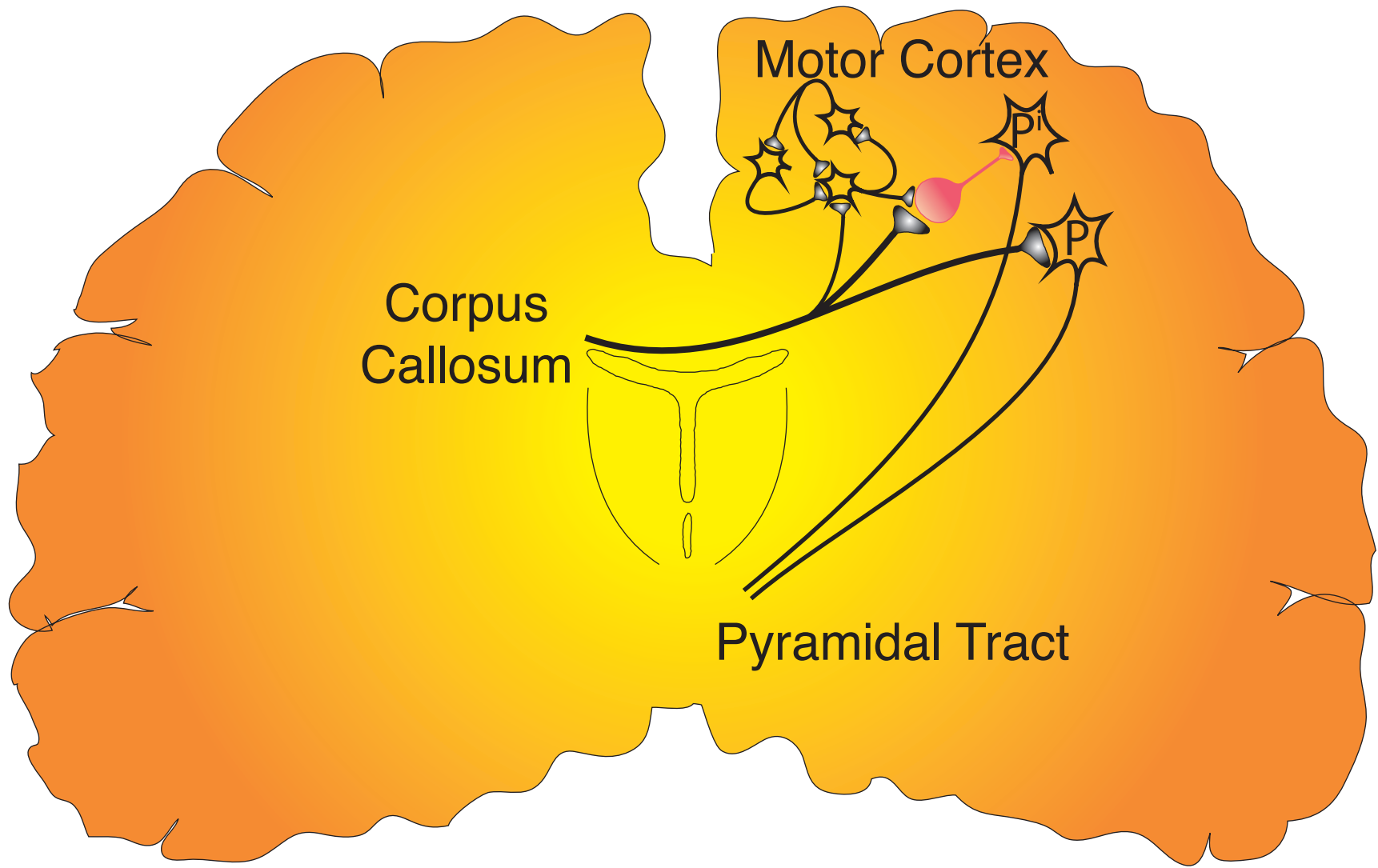

# Figure 2

A

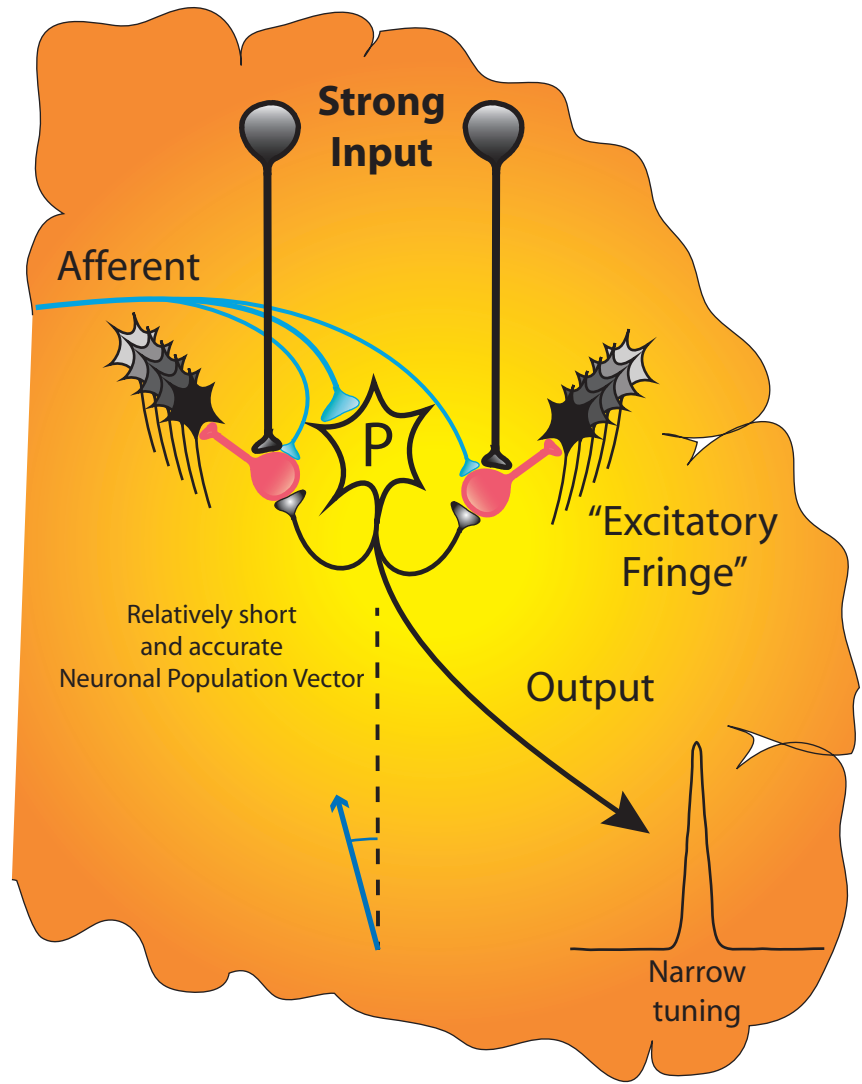

B

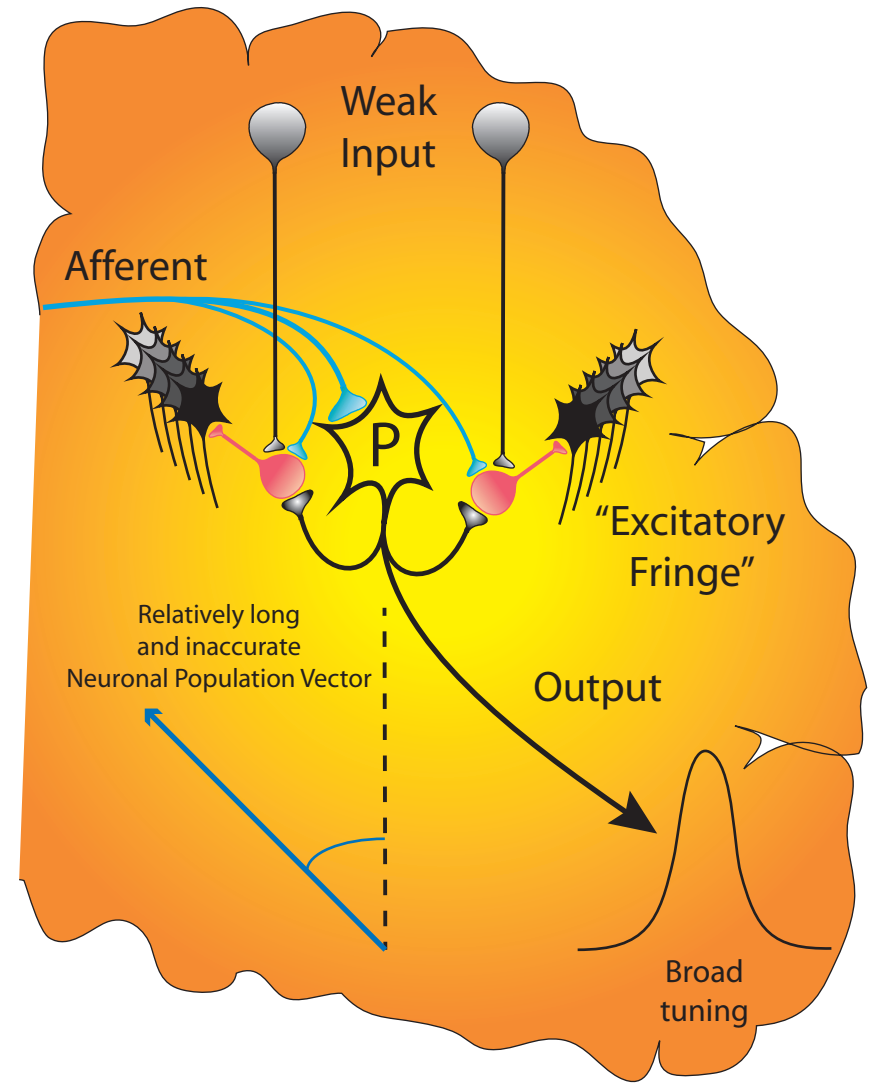

# Figure 3

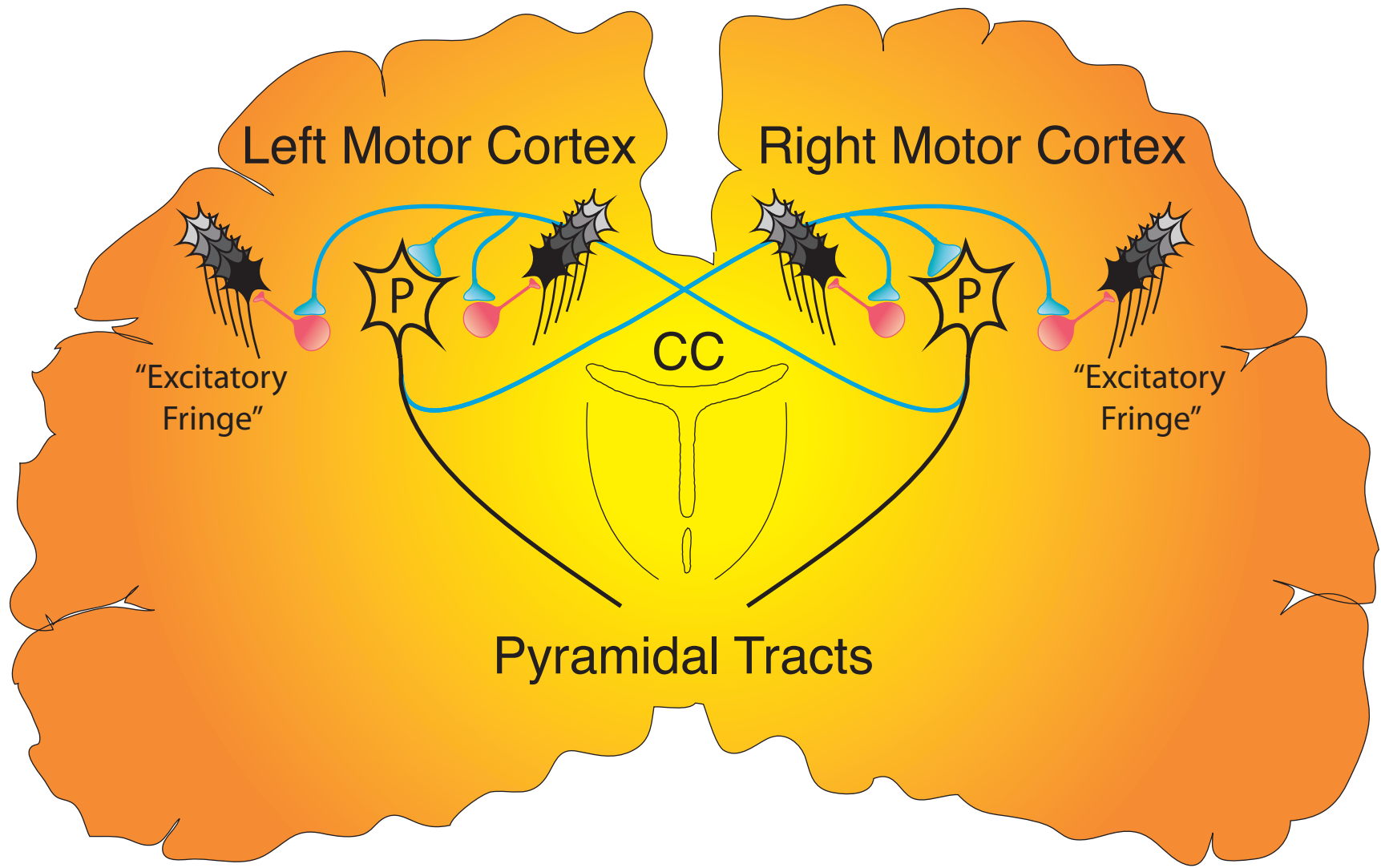

# Figure 4

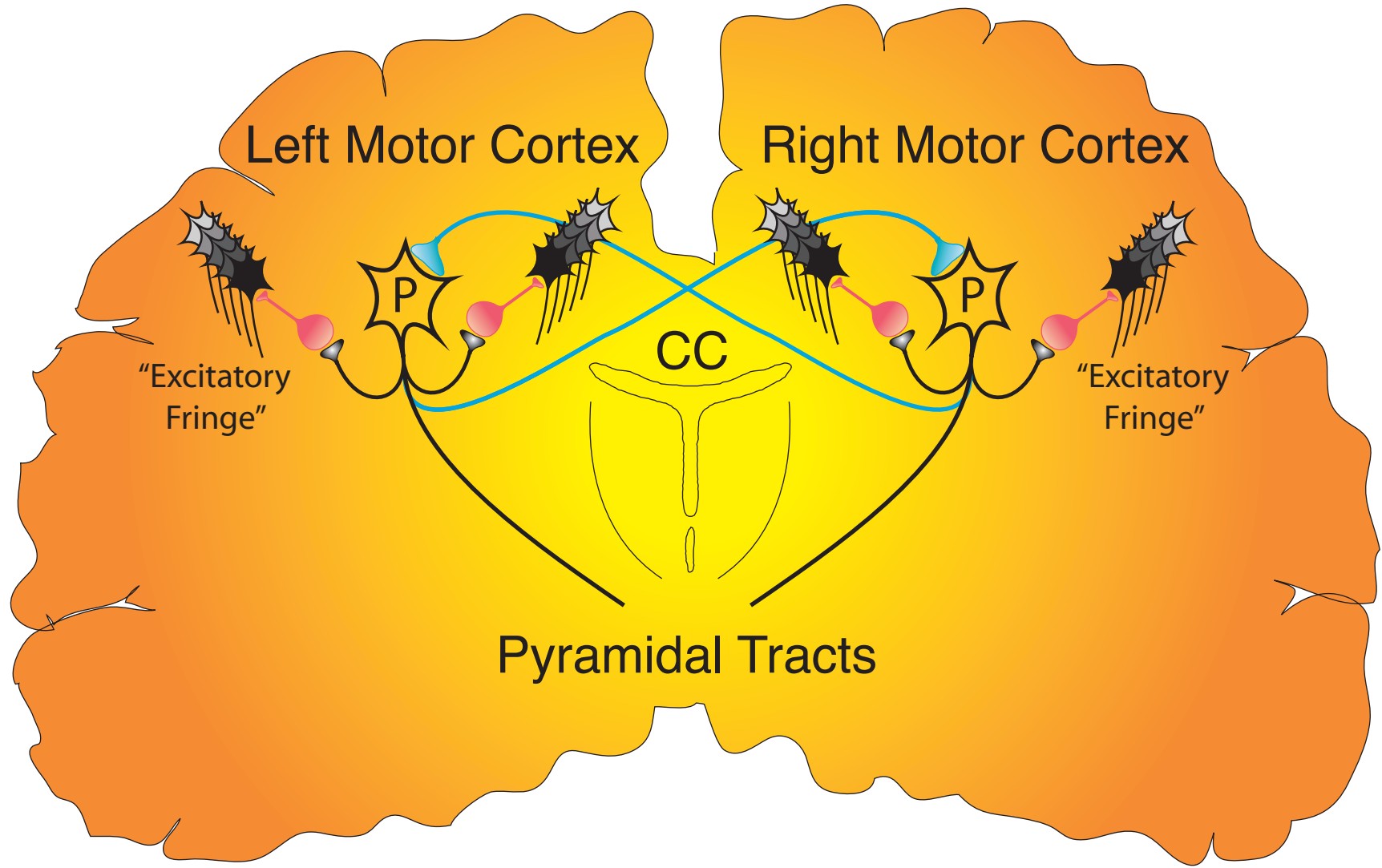